\documentclass[aps,prl,preprint,nopacs,superscriptaddress]{revtex4}
\usepackage{amsmath}
\usepackage{amssymb}
\usepackage{graphicx}
\usepackage{hyperref}
\newcommand{\beq}{\begin{equation*}}
\newcommand{\eeq}{\end{equation*}}

\newcommand{\mwt}{Mo$_x$W$_{1-x}$Te$_2$}

\begin{document}

\title{Topological Weyl phase transition in \mwt}

\author{Ilya Belopolski} \email{ilyab@princeton.edu}
\affiliation{Laboratory for Topological Quantum Matter and Spectroscopy (B7), Department of Physics, Princeton University, Princeton, New Jersey 08544, USA}

\author{Daniel S. Sanchez}
\affiliation{Laboratory for Topological Quantum Matter and Spectroscopy (B7), Department of Physics, Princeton University, Princeton, New Jersey 08544, USA}

\author{Yukiaki Ishida}
\affiliation{The Institute for Solid State Physics (ISSP), University of Tokyo, Kashiwa-no-ha, Kashiwa, Chiba 277-8581, Japan}

\author{Peng Yu}
\affiliation{Centre for Programmable Materials, School of Materials Science and Engineering, Nanyang Technological University, 639798, Singapore}

\author{Songtian S. Zhang}
\affiliation{Laboratory for Topological Quantum Matter and Spectroscopy (B7), Department of Physics, Princeton University, Princeton, New Jersey 08544, USA}

\author{Tay-Rong Chang}
\affiliation{Department of Physics, National Tsing Hua University, Hsinchu 30013, Taiwan}

\author{Guoqing Chang}
\affiliation{Centre for Advanced 2D Materials and Graphene Research Centre, National University of Singapore, 6 Science Drive 2, 117546, Singapore}
\affiliation{Department of Physics, National University of Singapore, 2 Science Drive 3, 117546, Singapore}

\author{Xingchen Pan}
\affiliation{National Laboratory of Solid State Microstructures, Collaborative Innovation Center of Advanced Microstructures, and Department of Physics, Nanjing University, Nanjing, 210093, China}

\author{Hong Lu}
\affiliation{International Center for Quantum Materials, Peking University, Beijing 100871, China}

\author{Hao Zheng}
\affiliation{Laboratory for Topological Quantum Matter and Spectroscopy (B7), Department of Physics, Princeton University, Princeton, New Jersey 08544, USA}

\author{Su-Yang Xu}
\affiliation{Laboratory for Topological Quantum Matter and Spectroscopy (B7), Department of Physics, Princeton University, Princeton, New Jersey 08544, USA}

\author{Baigeng Wang}
\affiliation{National Laboratory of Solid State Microstructures, Collaborative Innovation Center of Advanced Microstructures, and Department of Physics, Nanjing University, Nanjing, 210093, China}

\author{Guang Bian}
\affiliation{Laboratory for Topological Quantum Matter and Spectroscopy (B7), Department of Physics, Princeton University, Princeton, New Jersey 08544, USA}

\author{Da-Wei Fu}
\affiliation{Ordered Matter Science Research Center, College of Chemistry and Chemical Engineering, Southeast University, Nanjing 211189, China.}

\author{Shisheng Li}
\affiliation{Centre for Advanced 2D Materials and Graphene Research Centre, National University of Singapore, 6 Science Drive 2, 117546, Singapore} \affiliation{Department of Physics, National University of Singapore, 2 Science Drive 3, 117546, Singapore}

\author{Goki Eda}
\affiliation{Centre for Advanced 2D Materials and Graphene Research Centre, National University of Singapore, 6 Science Drive 2, 117546, Singapore} \affiliation{Department of Physics, National University of Singapore, 2 Science Drive 3, 117546, Singapore} \affiliation{Department of Chemistry, National University of Singapore, 3 Science Drive 3, 117543, Singapore}

\author{Horng-Tay Jeng}
\affiliation{Department of Physics, National Tsing Hua University, Hsinchu 30013, Taiwan}
\affiliation{Institute of Physics, Academia Sinica, Taipei 11529, Taiwan}

\author{Takeshi Kondo}
\affiliation{The Institute for Solid State Physics (ISSP), University of Tokyo, Kashiwa-no-ha, Kashiwa, Chiba 277-8581, Japan}

\author{Shuang Jia}
\affiliation{International Center for Quantum Materials, Peking University, Beijing 100871, China}

\author{Hsin Lin}
\affiliation{Centre for Advanced 2D Materials and Graphene Research Centre, National University of Singapore, 6 Science Drive 2, 117546, Singapore} \affiliation{Department of Physics, National University of Singapore, 2 Science Drive 3, 117546, Singapore}

\author{Zheng Liu}
\affiliation{Centre for Programmed Materials, School of Materials Science and Engineering, Nanyang Technological University, 639798, Singapore}
\affiliation{Centre for Micro/nanoelectronics (NOVITAS), School of Electrical and Electronic Engineering, Nanyang Technological University, 639798, Singapore}
\affiliation{CINTRA CNRS/NTU/THALES, UMI 3288, Research Techno Plaza, 50 Nanyang Drive, Border X Block, Level 6, 637553, Singapore}

\author{Fengqi Song}
\affiliation{National Laboratory of Solid State Microstructures, Collaborative Innovation Center of Advanced Microstructures, and Department of Physics, Nanjing University, Nanjing, 210093, China}

\author{Shik Shin}
\affiliation{The Institute for Solid State Physics (ISSP), University of Tokyo, Kashiwa-no-ha, Kashiwa, Chiba 277-8581, Japan}

\author{M. Zahid Hasan} \email{mzhasan@princeton.edu}
\affiliation{Laboratory for Topological Quantum Matter and Spectroscopy (B7), Department of Physics, Princeton University, Princeton, New Jersey 08544, USA}


\begin{abstract}
Topological phases of matter exhibit phase transitions between distinct topological classes. These phase transitions are exotic in that they do not fall within the traditional Ginzburg-Landau paradigm but are instead associated with changes in bulk topological invariants and associated topological surface states. In the case of a Weyl semimetal this phase transition is particularly unusual because it involves the creation of bulk chiral charges and the nucleation of topological Fermi arcs. Here we image a topological phase transition to a Weyl semimetal in \mwt\ with changing composition $x$. Using pump-probe ultrafast angle-resolved photoemission spectroscopy (pump-probe ARPES), we directly observe the nucleation of a topological Fermi arc at $x_c \sim 7\%$, showing the critical point of a topological Weyl phase transition. For Mo dopings $x < x_c$, we observe no Fermi arc, while for $x > x_c$, the Fermi arc gradually extends as the bulk Weyl points separate. Our results demonstrate for the first time the creation of magnetic monopoles in momentum space. Our work opens the way to manipulating chiral charge and topological Fermi arcs in Weyl semimetals for transport experiments and device applications.
\end{abstract}

\date{\today}
\maketitle

Topological phases of matter exhibit topological phase transitions (TPTs) which are remarkably different from phase transitions in the Ginzburg-Landau paradigm. Traditional Ginzburg-Landau phase transitions are typically associated with spontaneous symmetry breaking and the development of a local, continuous-valued order parameter. By contrast, TPTs do not break any symmetry and correspond to a change in a global topological index which typically takes on discrete values. Perhaps the most famous example of a TPT is the quantum Hall effect, observed in 1980, where the Hall conductivity transitions from one quantized plateau to the next under changing magnetic field \cite{QH}. More recently, a TPT was observed from a trivial insulator to a $\mathbb{Z}_2$ topological insulator in BiTl(S$_{1-\delta}$Se$_{\delta}$)$_2$ \cite{TI_Suyang_2011}. In that experiment, angle-resolved photoemission spectroscopy (ARPES) provided a striking example of a bulk band gap closing and re-opening as a function of $\delta$, with direct observation of a gapless Dirac cone surface state crossing the bulk band gap in the topological phase. All TPTs so far, including these two examples, have been observed as a phase transition from one bulk insulating phase to another. However, the discovery of the first Weyl semimetal has set the stage for the first TPT to a gapless phase \cite{TaAsUs, LingLu, TaAsThem, TaAsThyUs, TaAsThyThem, NbAs, TaPUs}. Specifically, for a TPT to a Weyl semimetal, the bulk band gap closes and then splits into Weyl points. This phase transition is fascinating because the nucleation of Weyl points corresponds to the creation of chiral charge, or equivalently, the creation of magnetic monopoles in momentum space \cite{Murakami, Multilayer, Pyrochlore, Vish}. This transition is also associated with the development of a topological Fermi arc surface state which similarly nucleates from the bulk band gap closing and connects the Weyl points in the surface Brillouin zone, extending across the surface as the Weyl points move apart in the bulk. There may further arise unusual behavior at the critical point, where in generic cases the bulk band touching does not occur at a time-reversal invariant momentum and where the form of the dispersion at the critical degeneracy point is constrained by the requirement that it carry no net chiral charge. Despite these fascinating properties, to date a topological phase transition to a Weyl semimetal has yet to be realized.

Recently, it was predicted in \textit{ab initio} calculation that \mwt\ is a tunable Weyl semimetal \cite{TR}. It was concurrently predicted that the end compound WTe$_2$ hosts a novel type of strongly Lorentz-violating, or Type II, Weyl fermion \cite{Andrei_WT_2015}, with related predictions following for MoTe$_2$ \cite{Zhijun_MT_2015, Binghai_MT_2015}. However, it became clear that both of these end compounds lie close to a topological phase transition, so that calculation could not conclusively predict the topological phase. Considerable experimental efforts then provided evidence for a Weyl semimetal in MoTe$_2$, although no evidence or discussion was presented regarding the topological phase transition in \mwt\ \cite{myotherothermowte, myothermowte, HaoSTM, Adam1, Shuyun, Baumberger}. Here we use pump-probe angle-resolved photoemission spectroscopy (pump-probe ARPES) to carry out a systematic study of the \mwt\ series from $x = 0\%$ to $50\%$. We directly access the topological Fermi arc above the Fermi level. We find that undoped WTe$_2$ is topologically trivial and we directly observe a topological phase transition from a topologically trivial phase to a Weyl semimetal in \mwt\ by changing $x$, with $x_c \sim 7\%$. We perform novel \textit{ab initio} calculations to pinpoint the topological phase transition in numerics and we find excellent agreement with our experimental results. We directly demonstrate the first topological phase transition between a trivial semimetal and a Weyl semimetal.

We first provide an overview of the crystal and electronic structure of \mwt. The creation of chiral charge can be understood as a process of gap closing or band inversion, Fig. \ref{Fig1}a-c. The crystal structure of \mwt\ consists of stacked trilayers of Mo/W and Te, Fig. \ref{Fig1}d, e. The system is gapped throughout the bulk Brillouin zone, except near $\Gamma$, where the bulk valence and conduction bands overlap, Fig. \ref{Fig1}f. The crystal structure breaks inversion symmetry, so generically we expect Weyl points to arise where the bands overlap. A detailed \textit{ab initio} calculation shows that for most $x$, \mwt\ hosts 8 Weyl points near $\Gamma$ which lie on $k_z = 0$, at $k_y \sim \pm k_W$ and at energies $< 0.1$ eV above $\varepsilon_F$, Fig. \ref{Fig1}g, h \cite{TR}. Crucially, calculation predicts that as $x$ varies, the Weyl points move: for $x < x_{c}$ the system is in a trivial phase without Weyl points; at $x_{c}$ the bands intersect and nucleate Weyl points in four pairs of $W_1$ and $W_2$, giving a phase with 8 Weyl points; for $x_{c} < x < x'_{c}$ the $W_1$ and $W_2$ separate out; at $x = x'_{c}$ the $W_1$ meet up at $k_x = 0$ and annihilate; leaving the system in a phase with 4 Weyl points for $x > x'_{c}$. In the 8 Weyl point phase, topological theory and calculation show that there are topological Fermi arc surface states which connect each adjacent pair of $W_1$ and $W_2$ in the surface Brillouin zone, Fig. \ref{Fig1}i.

Next, we show a topological phase transition in \mwt. In pump-probe ARPES, we use a $1.48$ eV pump laser to first excite electrons into unoccupied electron states and then we use a $5.92$ eV probe laser to perform photoemission on the excited electrons, granting direct access in experiment to the unoccupied band structure. We present an $E_\textrm{B}$-$k_x$ spectrum at $k_y \sim k_W$ for $x = 0\%, 7\%, 20\%, 25\%, 40\%, 50\%$, Fig. \ref{Fig2}a-f. For compositions $x \geq 20\%$ we observe a short, bright band extending above $\varepsilon_F$ (purple curve), with dimmer bands on either side (green curve). We also observe slight kinks where the bright and dim bands meet. We also see the bright band grow in length systematically with $x$. We interpret the bright bands as disjoint Fermi arcs connecting Weyl points. This interpretation is consistent with the locations of Weyl points in calculation. Also, the high spectral weight suggests a surface state well-localized at the surface, while the dimmer states suggest hybridization with the bulk. Lastly, the increase in length of the Fermi arc with $x$ is consistent with increasing separation of the Weyl points for heavier Mo doping. At $x = 7\%$, we see no Fermi arc, but find a bright spot (purple arrow) and at $x = 0\%$ we find a large smooth band without any particular features (green curve). We interpret the bright spot as the nucleation of a topological Fermi arc, signalling that at $x_c = 7\%$ the system is at the critical point for a topological phase transition. We interpret the large band at $x = 0\%$ as a signature of a topologically trivial phase in undoped WTe$_2$. We find that this evolution of the band structure is consistent and robust regardless of whether samples were grown by a flux technique or chemical vapor transport (CVT); regardless of whether I$_2$ or TeI$_4$ was used as the transport agent in CVT growth; and despite differences in heating temperatures and durations, see the Methods section in the Supplementary Materials for additional details. We further interpret our results as the Weyl points ``snipping'' a Fermi arc out of a trivial surface state. WTe$_2$ is in a trivial phase, with a large trivial electron-like surface state, Fig. \ref{Fig2}g. At the critical point, the bulk band gap closes on or near the trivial surface state, Fig. \ref{Fig2}h. Then, in the Weyl semimetal phase, the Weyl points separate out, threading a Fermi arc and erasing the trivial surface state \ref{Fig2}i, j. Although only the topological Fermi arc terminates strictly on the Weyl points, the trivial surface state still merges into the bulk very close to the Weyl points, so that within the linewidth and resolution of our measurement, we still see that the topological and trivial arc connect. However, we observe that they meet at slightly different slopes, forming a kink in the surface state bands which demonstrates a Weyl semimetal. By directly observing the nucleation of a topological Fermi arc and its systematic evolution with increasing Mo doping $x$, we demonstrate a topological phase transition from a trivial phase to a Weyl semimetal in \mwt.

We perform novel \textit{ab initio} calculations on the \mwt\ series and for the first time we pinpoint the topological phase transition in numerics. We find that our calculations directly support our experimental results. First, we find excellent agreement in WTe$_2$ between the band structure in pump-probe ARPES and calculation both above and below the Fermi level, Fig. \ref{Fig2}a and Fig. \ref{Fig3}a. Specifically, we find two electron-like pockets above $E_\textrm{F}$, a hole pocket below $E_\textrm{F}$ as well as the broad hole-like bulk continuum extending to $\sim 0.05$ eV above $E_\textrm{F}$, all consistent between experiment and theory. We also find excellent agreement for other compositions $x$, for other measurement photon energies, and for constant energy $k_x$-$k_y$ surfaces as well as for $E_\textrm{B}$-$k_x$ cuts, see Ref. \cite{myothermowte, myotherothermowte}. We also find that both the energy and momentum separation of the $W_1$ and $W_2$ increase with $x$ in both experiment and theory, Figs \ref{Fig3}b, c. Next, we zoom in near the Weyl points (red rectangle in \ref{Fig3}a) and study the band structure with increasing Mo, Fig. \ref{Fig3}d. For WTe$_2$, we see that the bulk valence and conduction bands have an indirect gap. At very small Mo doping, $x < 1\%$, the bands touch, and for larger $x$, the bands invert and give rise to two Weyl points. Consistent with pump-probe ARPES, while WTe$_2$ hosts a large trivial surface state, for larger $x$ the Weyl points snip out a topological Fermi arc. We find that calculation reproduces all aspects of our pump-probe ARPES data, but likely underestimates the size of the trivial gap in WTe$_2$ because the topological phase transition occurs around $x \sim 7\%$ in experiment while it occurs for exceedingly small $x$ in calculation. At the same time, the calculation systematically underestimates the separation of the Weyl points at $x > 20\%$. This is perhaps reasonable because the calculation for arbitrary doping $x$ uses an interpolation of Wannier models which is expected to be most accurate for small deviations from stoichiometric WTe$_2$ \cite{TR}.

Next, we study systematic pump-probe ARPES spectra to provide further evidence for a topological phase transition in \mwt. Specifically, we study $E_\textrm{B}$-$k_x$ cuts of $x = 0\%, 7\%, 20\%$ at different $k_y$, Fig. \ref{Fig4}a-c. For stoichiometric WTe$_2$, our data shows that the trivial surface state is unbroken and no signature of a Fermi arc or critical point appears at any $k_y$, again confirming that WTe$_2$ is in a robust trivial phase. At $x = 7\%$, we can identify the critical point (purple arrow) in the spectra with a range of $k_y$. At the same time, we can identify a single $k_y \sim k_W$, center panel of Fig. \ref{Fig4}b, where the critical point is brightest, as expected since the bulk bands strictly meet at a single point in momentum space. Away from the special value of $k_y = k_W$, the critical point broadens and fades. Lastly, at $x = 20\%$, we can directly follow the dispersion of the Fermi arc in $k_y$ (endpoints marked by red and blue arrows). In this way, we confirm that the length of the arc remains consistent as we sweep through $k_y$. We show analogous systematics for all other compositions in the Supplementary Materials. We also use these results to measure the Fermi velocity of the Weyl cones along certain directions by mapping out the attachment curves of the Fermi arcs to the bulk bands, see also the Supplementary Materials. Our systematic pump-probe ARPES spectra confirm that we have observed a topological phase transition in \mwt.

We summarize our results on a phase diagram for \mwt, Fig. \ref{Fig5}a. We discover that the \mwt\ series includes a trivial phase under low doping $x < x_c \sim 7\%$ and a phase with 8 Weyl points beyond $x_c$. While we include MoTe$_2$ ($x = 100\%$) in our phase diagram, we emphasize that in calculation MoTe$_2$ is also close to a critical point, like WTe$_2$. Recent works predict either a Weyl semimetal with 4 or 8 Weyl points in MoTe$_2$, as indicated by the hashed box around $c_2$ in Fig. \ref{Fig5}a, although recent experimental works appear to lean toward the 8 Weyl point case. Next, we can place our result in perspective by comparing it with the phase transition to a topological insulator, Fig. \ref{Fig5}b, c. In the same way that the transition to a topological insulator threads a Dirac cone surface state through the bulk band gap, here we observe how a transition to a Weyl semimetal threads a topological Fermi arc surface state across the surface Brillouin zone. By directly observing the nucleation of a topological Fermi arc, we have demonstrated the first topological phase transition to a Weyl semimetal. Our discovery also provides the first demonstration of the nucleation of chiral charge in a real material. We note that the tunability demonstrated here has never been observed in a Weyl semimetal and may be relevant to applications, especially since our results indicate that certain Weyl points sit very close to the Fermi level in \mwt\ for a wide range of $x$. Our work represents a fundamental advance in the study of topological phases and presents new potential for applications based on a Weyl semimetal tunable through the topological phase transition.

\section{Acknowledgements}

I.B. acknowledges the support of the US National Science Foundation GRFP. Work at Princeton University is supported by the Emergent Phenomena in Quantum Systems (EPiQS) Initiative of the Gordon and Betty Moore Foundation under Grant No. GBMF4547 (M.Z.H.) and by the National Science Foundation, Division of Materials Research, under Grants No. NSF-DMR-1507585 and No. NSF-DMR-1006492. Y.I. is supported by the Japan Society for the Promotion of Science, KAKENHI 26800165. T.-R.C. and H.-T.J. were supported by the National Science Council, Taiwan. H.-T.J. also thanks the National Center for High-Performance Computing, Computer and Information Network Center National Taiwan University, and National Center for Theoretical Sciences, Taiwan, for technical support. X.C.P., Y.S., H.J.B., G.H.W and F.Q.S. thank the National Key Projects for Basic Research of China (Grant Nos. 2013CB922100, 2011CB922103), the National Natural Science Foundation of China (Grant Nos. 91421109, 11522432, and 21571097) and the NSF of Jiangsu province (No. BK20130054). This work is also financially supported by the Singapore National Research Foundation under NRF Award No. NRF-NRFF2013-03 (H.L.); NRF Research Fellowship (NRF-NRFF2011-02) and the NRF Medium-sized Centre program (G.E.); and NRF RF Award No. NRF-RF2013-08, the start-up funding from Nanyang Technological University (M4081137.070) (Z.L.).

\clearpage
\begin{figure*}
\centering
\includegraphics[width=16cm,trim={1in 1in 1in 1in},clip]{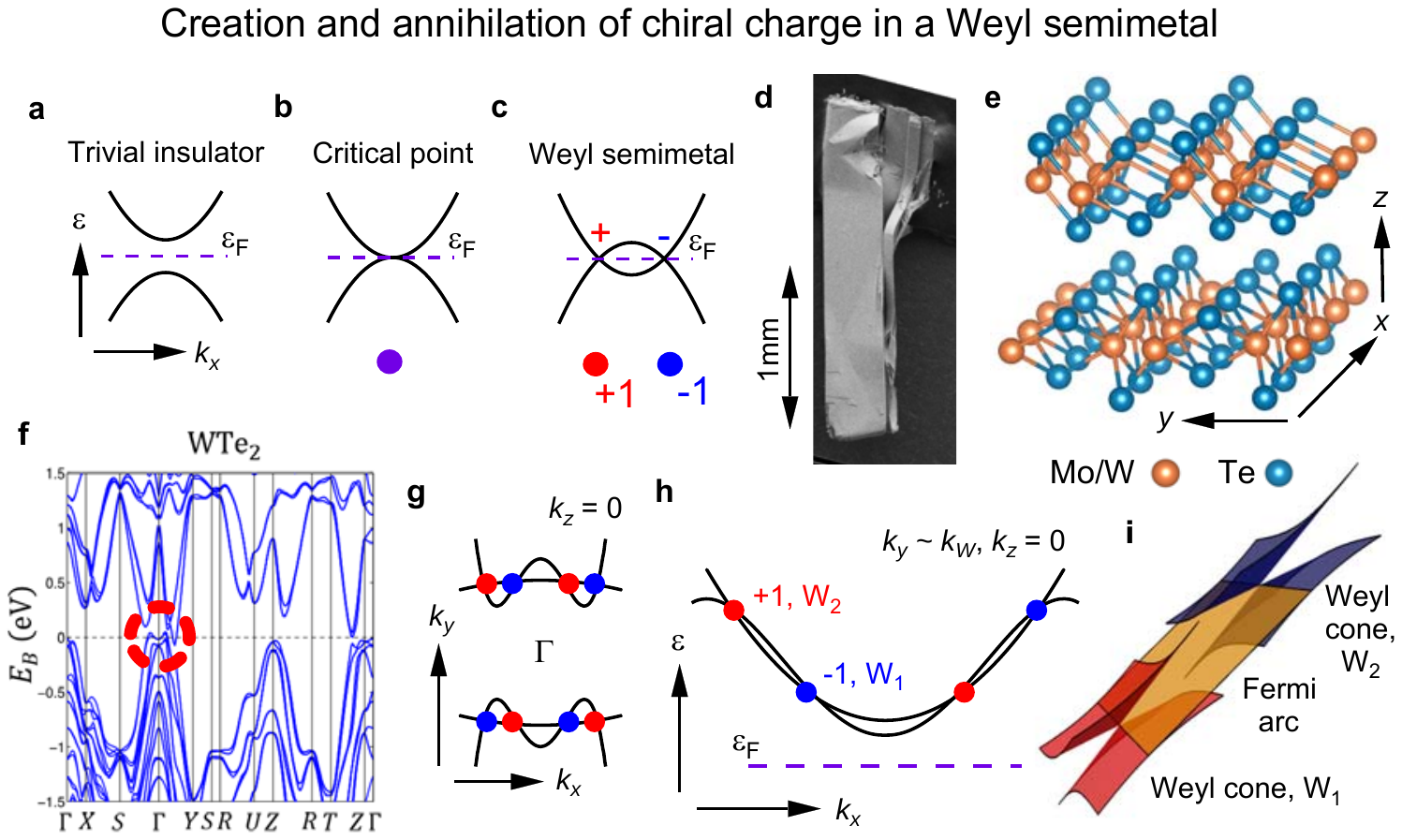}
\end{figure*}

\clearpage
\begin{figure}
\caption{\label{Fig1} \textbf{Creation of Weyl points in a topological phase transition.} {\bf a}-{\bf c}, Schematic of a band inversion which gives rise to Weyl points. Initially the system is gapped, (a). As the bands invert, they first touch at a point, (b). This point can be viewed as a band crossing consisting of two chiral charges on top of each other in momentum space (purple dot). Upon further inversion, the chiral charges separate, giving a Weyl semimetal (c). {\bf d}, High-resolution scanning electron microscope (SEM) image of a single crystal of \mwt, showing the layered structure. {\bf e}, Crystal structure of \mwt. {\bf f}, Overview of the band structure of \mwt, showing a band crossing near $\Gamma$ (red circle). {\bf g}, {\bf h}, Schematic of the configuration of Weyl points. All Weyl points are on $k_z = 0$, but at different energies, as shown \cite{TR}. {\bf i}, The Weyl points are Type II, meaning the Weyl cones are tipped over on their side and connected as illustrated by a Fermi arc.}
\end{figure}

\clearpage
\begin{figure}
\centering
\includegraphics[width=16cm,trim={1in 1in 1in 1in},clip]{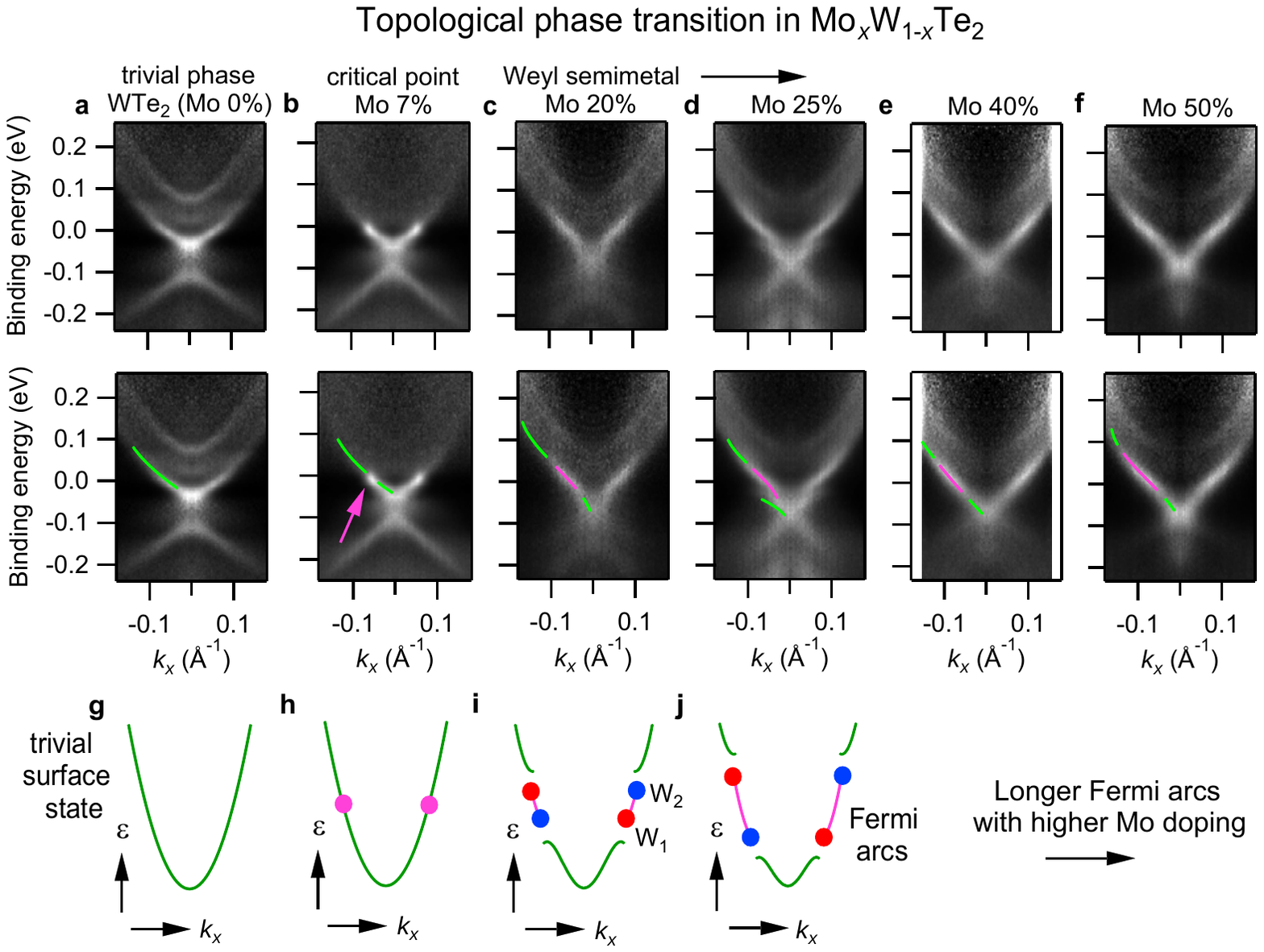}
\end{figure}

\clearpage
\begin{figure}
\caption{\label{Fig2} \textbf{Topological phase transition in \mwt.} {\bf a}-{\bf f}, Evolution of \mwt\ with $x$ in experiment. The top row shows raw data, the second row shows the same data with additional hand-drawn markings. For undoped WTe$_2$, $x = 0\%$, we see an unbroken trivial surface state (green curve) indicating a trivial phase, (a). At small doping, we find a critical point (purple arrow), (b). For larger dopings, we observe a topological Fermi arc (purple curve), which systematically grows with $x$, (c-f). {\bf g}-{\bf j}, Schematic of the experimental results in (a-f). We can view the Weyl points as ``snipping'' out a topological Fermi arc from the large trivial surface state.}
\end{figure}

\clearpage
\begin{figure}
\centering
\includegraphics[width=16cm,trim={1in 1in 1in 1in},clip]{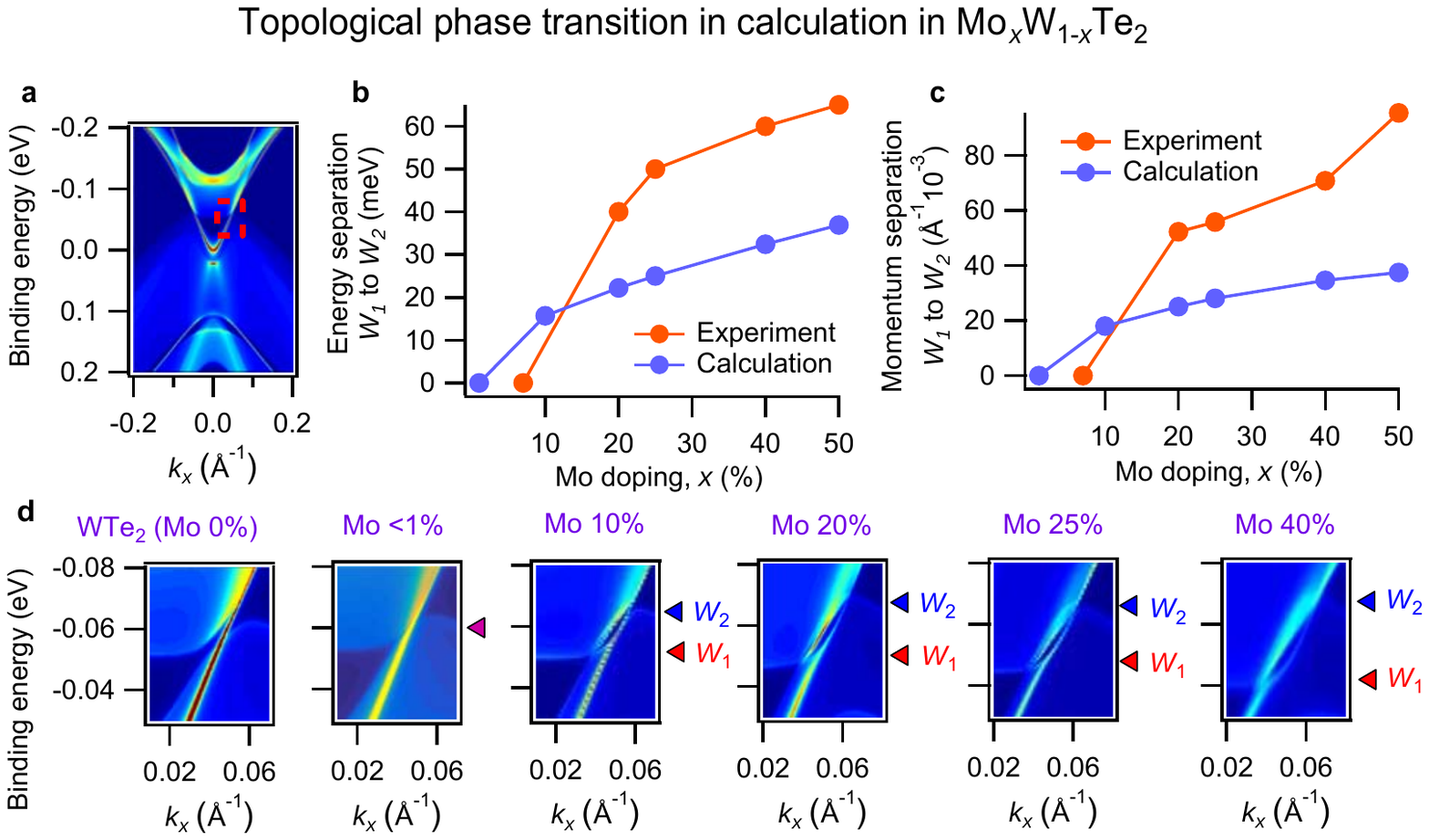}
\end{figure}

\clearpage
\begin{figure}
\caption{\label{Fig3} \textbf{Topological phase transition in calculation.} {\bf a}, Calculation for undoped WTe$_2$, showing excellent agreement with our ARPES data. {\bf b}, {\bf c}, Calculated energy and momentum separation of the $W_1$ and $W_2$ compared with experiment. {\bf d}, Calculation near $W_1$ and $W_2$ at different doping $x$, in the region marked by the red dotted square in (a). We observe the same trend in calculation and experiment, but we find that the trivial phase is more robust in experiment and the calculation underestimates the separation of the Weyl points for large $x$.}
\end{figure}

\clearpage
\begin{figure}
\centering
\includegraphics[width=16cm,trim={1in 1in 1in 1in},clip]{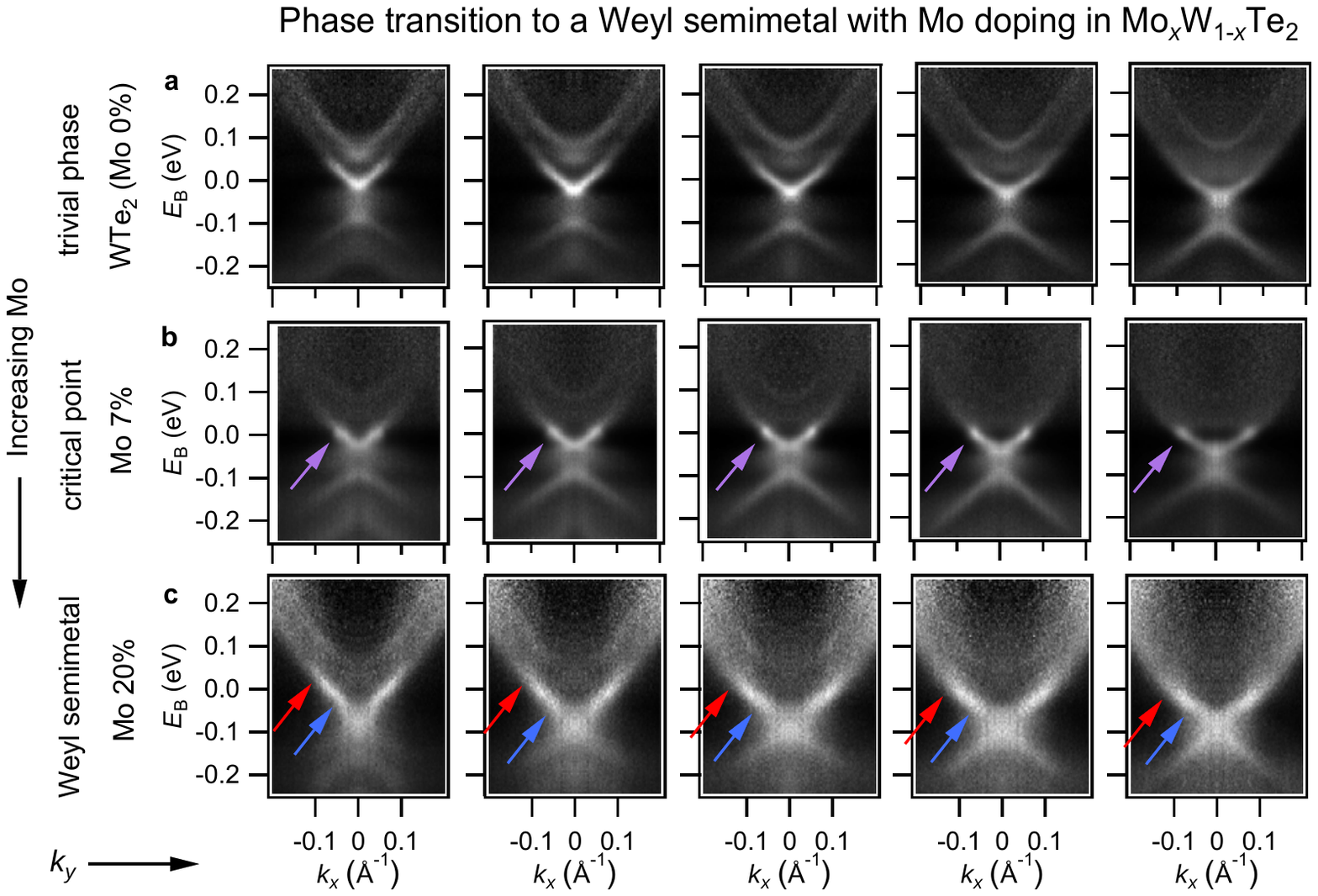}
\end{figure}

\clearpage
\begin{figure}
\caption{\label{Fig4} \textbf{Systematic study of the topological phase transition in \mwt.} {\bf a}-{\bf c}, ARPES spectra at different $k_y$ for different compositions $x$. We see that at $x = 0\%$ for all $k_y$ the trivial state is smooth, (a). At the critical doping, the band touching (purple arrow) is consistent across a narrow range of $k_y$, (b). We can identify the Weyl cone projections by looking at the ends of the Fermi arc (red and blue arrows). We can follow the dispersion of the arc in $k_y$ and use it to indirectly measure the Fermi velocity of the Weyl cones using only the observed surface state band structure, (c), see also the Supplementary Materials.}
\end{figure}

\clearpage
\begin{figure}
\centering
\includegraphics[width=16cm,trim={1in 1in 1in 1in},clip]{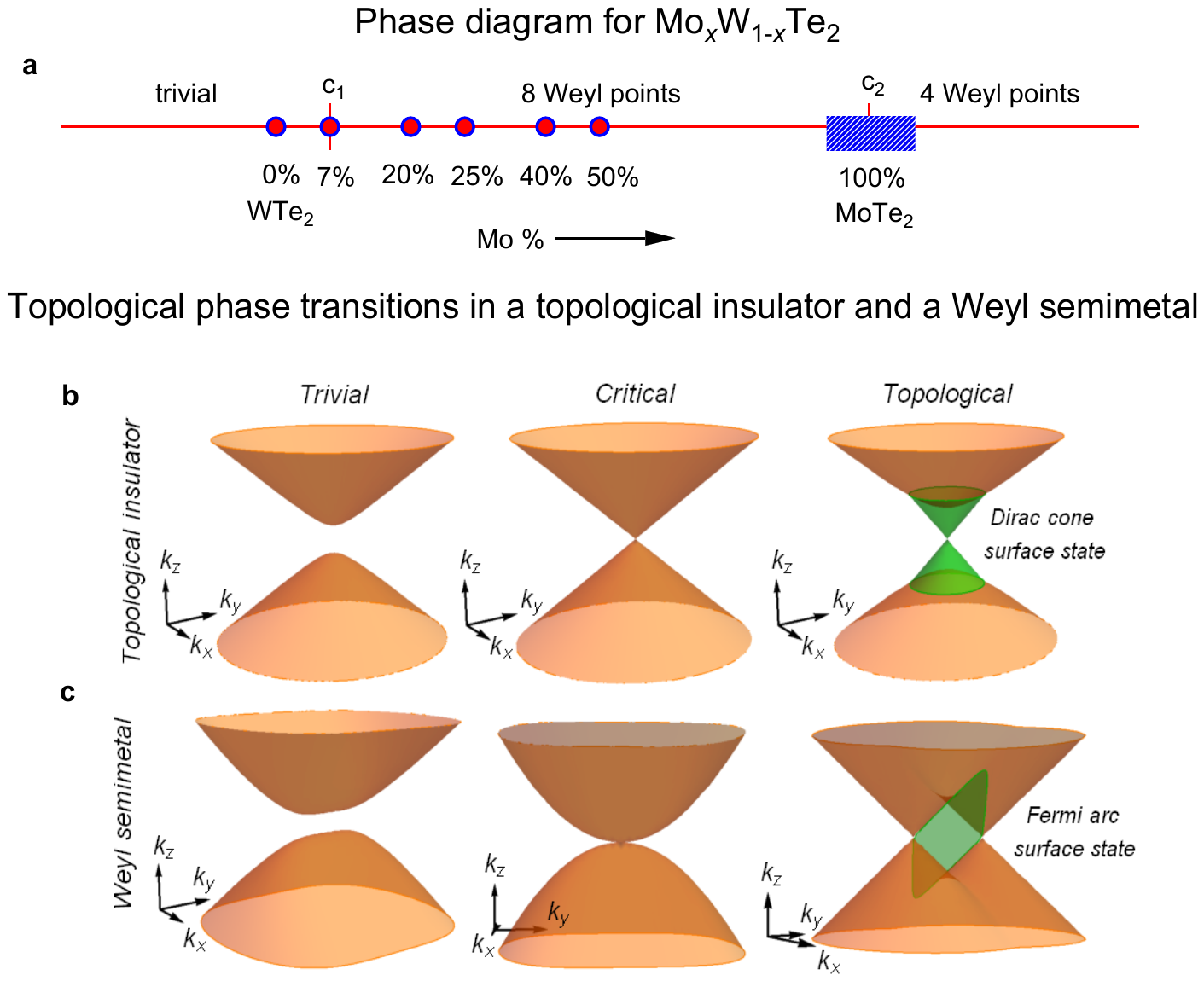}
\end{figure}

\clearpage
\begin{figure}
\caption{\label{Fig5} \textbf{Creation of chiral charge in \mwt.} {\bf a}, Phase diagram for \mwt. We observe a phase transition from a trivial phase to a Weyl semimetal phase with eight Weyl points, across the critical point $c_1$.  There may also be a second critical point to a four Weyl point phase, $c_2$, for $x \sim 100\%$, although experimental studies suggest that stoichiometric MoTe$_2$ remains in an eight Weyl point phase. {\bf b}, {\bf c}, Comparison of the phase transition to a topological insulator and a Weyl semimetal. The topological insulator has a gapless critical point, while the Weyl semimetal passes to a gapless critical phase. In the simplest case, the topological insulator has a bulk linear Dirac dispersion at the critical point, while the critical point dispersion in a Weyl semimetal cannot be linear in all three momentum-space directions because it cannot carry chiral charge. This may give rise to unusual transport properties at the critical point for a topological phase transition in \mwt.}
\end{figure}

\clearpage

\bigskip
\begin{center}
\textbf{\large Supplementary Materials}
\end{center}
\bigskip

\setcounter{equation}{0}
\setcounter{figure}{0}
\setcounter{table}{0}
\makeatletter
\renewcommand{\theequation}{S\arabic{equation}}
\renewcommand{\thefigure}{S\arabic{figure}}
\renewcommand{\thetable}{S\arabic{table}}

\noindent In these Supplementary Materials, we first provide additional systematic experimental data for \mwt\ for higher dopings $x$. Then, we present an interesting application of our results to the measurement of the Fermi velocity of the bulk Weyl cones of \mwt. Lastly, we provide our methods.

\section{Systematic data at high doping}

We present additional systematic measurements of \mwt\ for high dopings, $x = 25\%$, $40\%$, $50\%$, see Fig. \ref{Fig1}a-c. These spectra complement Fig. 4 of the main text. We can track the Fermi arc systematically for a wide range of $k_y$. We further find that the Fermi arc disperses into the unoccupied side for larger $k_y$. Again, the arc becomes longer with increasing Mo doping and we can observe this evolution consistently by studying the dependence on Mo doping for any $k_y$. Our results are consistent with topologial theory, \textit{ab initio} calculation and our pump-probe ARPES spectra presented in the main text. Our systematic data on \mwt\ with higher Mo doping supports our observation of a topological phase transition.

\section{Fermi velocity of the Weyl cones in the bulk}

We present a separate, neat consequence of our results. Using our pump-probe ARPES data, we can indirectly estimate the Fermi velocity of the Weyl cone using only the surface state band structure. In particular, we map out the termination of the Fermi arc on spectra taken at a range of $k_y$. Then, we can determine the Fermi velocity of the bulk Weyl cone along the attachment direction of the topological Fermi arc, as illustrated in Fig. \ref{Fig2}a. We perform this analysis for $x = 25\%$ and $50\%$, Fig. \ref{Fig2}b, c. We perform a linear fit of the result to find attachment Fermi velocities. For the case of $x = 50\%$, we measure experimentally a velocity of $0.47\ \textrm{eV\AA}$ for $W_1$ and $0.59\ \textrm{eV\AA}$ for $W_2$. We further find an attachment angle of $\sim -4^{\circ}$ and $\sim 11^{\circ}$ for the attachment to $W_1$ and $W_2$, so that the Fermi arc approximately connects to the bulk along $k_y$, consistent with the approximate alignment of the Weyl points along $k_x$. In this way, we access properties of the bulk Weyl cones even with a surface-sensitive photoemission measurement. This approach is also fundamentally related to the topological invariants characterizing a Weyl semimetal in the sense that we take advantage of the termination of the surface state on the bulk in order to track the dispersion of the bulk band.

\section{Methods}

\subsection{Pump-probe ARPES}

Pump-probe ARPES measurements were carried out using a hemispherical Scienta R4000 analyzer and a mode-locked Ti:Sapphire laser system that delivered $1.48$ eV pump and $5.92$ eV probe pulses at a repetition rate of $250$ kHz \cite{IshidaMethods}. The time and energy resolution were $300$ fs and $15$ meV, respectively. The spot diameters of the pump and probe lasers at the sample were $250\ \mu$m and $85\ \mu$m, respectively. Measurements were carried out at pressures $<5 \times 10^{-11}$ Torr and temperatures $\sim8$ K.

\subsection{Crystal growth \& characterization}

\textbf{Mo doping $x = 0\%$ and $20\%$}: Large, well-formed, ribbon-like single crystals of \mwt\ for $x = 0\%$ and $20\%$ were grown by chemical vapor transport (CVT) with iodine (I) as the transporting gas. Prior to crystal growth, thermal and ultrasonic cleaning treatments of the quartz tubes were performed to avoid any contamination from the quartz container. Stoichiometric amounts of tungsten (W) powder (99.9\%, Sigma-Aldrich), molybdenum (Mo) powder (99.95\%, Sigma-Aldrich) and tellurium (Te) powder (99.95\%, Sigma-Aldrich) with a total weight of 500 mg, plus an extra 35 mg of I$_2$ as the transport gas were sealed in an evacuated $20$ cm long quartz tube under vacuum at $10^{-6}$ Torr. The quartz tube was placed in a three-zone furnance. First, the reaction zone was maintained at $850^{\circ}$C for 40 h with the growth zone at $900^{\circ}$C in order to prevent the transport of the product and a complete reaction; the reaction zone was then heated to $1070^{\circ}$C and held for 7 days with the growth zone at $950^{\circ}$C. Finally, the furnace was naturally cooled down to room temperature and the single crystals were collected in the growth zone. Residual I was cleaned using acetone and ethanol before measurement.\\

\textbf{Mo doping $x = 7\%$, $25\%$, $50\%$}: Single crystals of \mwt\ at $x = 7\%$ were grown via a self-flux technique. Mo, W and Te were mixed in a molar ratio of 7:93:5000, where Te excess acts as a flux. The mixture was heated to $1050^{\circ}$C for three days, and slowly cooled to $850^{\circ}$C over one week, followed by centrifugation to separate the crystal from the flux. Single crystals at $x = 25\%$ and $50\%$ were grown by iodine transport methods. A mix of stoichiometric Mo, W, Te and I$_2$ was sealed in a quartz tube and placed in a two-zone horizontal temperature gradient furnace. The high temperature side was kept at $1050^{\circ}$C for one week and the lower side at $950^{\circ}$C. Samples were characterized by an X-ray diffraction technique. A single crystal was placed on top of a glass fiber. Single-crystal X-ray diffraction measurements were carried out on a Rigaku Saturn 724 CCD based on a diffractometer operating at room temperature. The measurement was performed with Mo K$\alpha$ radiation, $\lambda = 0.71073\textrm{\AA}$, using the $\omega$-scan technique. The structure of the compound was determined by the direct methods routines in the SHELXS program and refined by full-matrix least squares methods on F2 in SHELXL.\\

\textbf{Mo doping $x = 40\%$}: Stoichiometric Mo, W and Te (2:8:10) powder was heated at $750^{\circ}$C for 4 days to synthesize polycrystalline \mwt. Single crystals were grown using CVT technique using TeI$_4$ as the transfer agent with a mass ratio 10:1. The quartz tube was sealed under vacuum and placed in a two-zone furnace. The temperatures of the two zones were maintained at $1000^{\circ}$C and $900^{\circ}$C for 2 weeks. The dopant distribution is not uniform, particularly near the crystal surface. The composition of the sample was determined by energy dispersive spectroscopy (EDS) using a scanning electron microscope (SEM).

\subsection{\textit{Ab initio} calculation}

The \textit{ab initio} calculations were based on the generalized gradient approximation (GGA) \cite{GGA} and used the full-potential projected augmented wave method \cite{PAW1,PAW2} as implemented in the VASP package \cite{PlaneWaves1}. Experimental lattice constants were used for both WTe$_2$ \cite{Bonding} and MoTe$_2$. A $15 \times 11 \times 7$ Monkhorst-Pack $k$-point mesh was used in the computations. The spin-orbit coupling effects were included in calculations. To calculate the bulk and surface electronic structures, we constructed a first-principles tight-binding model Hamiltonian by projecting onto the Wannier orbitals \cite{MLWF1,MLWF2,Wannier90}, using the VASP2WANNIER90 interface \cite{MLWF3}. We used W $d$ orbitals, Mo $d$ orbitals, and Te $p$ orbitals to construct Wannier functions, without performing the procedure for maximizing localization. The electronic structure of the \mwt\ samples with finite doping was calculated by a linear interpolation of the tight-binding model matrix elements of WTe$_2$ and MoTe$_2$. The surface states were calculated from the surface Green's function of the semi-infinite system \cite{Green}.

\begin{figure*}[h]
\centering
\includegraphics[width=16cm,trim={1in 1in 1in 1in},clip]{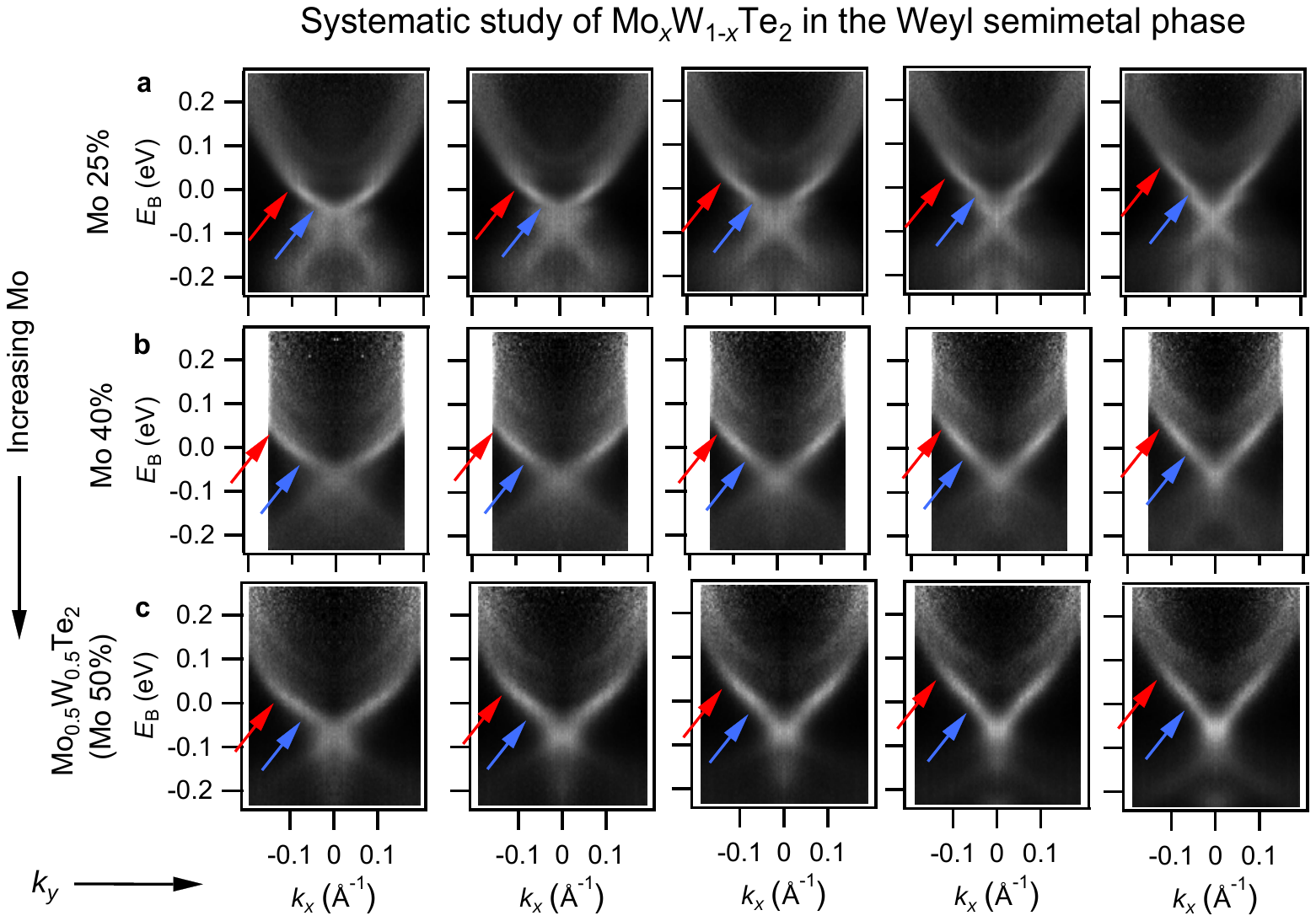}
\end{figure*}

\clearpage
\begin{figure*}
\centering
\caption{\label{Fig1} \textbf{Systematic study of \mwt\ at higher Mo doping.} \textbf{a}-\textbf{c}, Pump-probe ARPES spectra cutting at different $k_y$ for $x = 25\%$, $40\%$, $50\%$, complementing Fig. 4 in the main text, which shows analogous cuts for $x = 0\%$, $7\%$, $20\%$. The terminations of the Fermi arcs are marked by red and blue arrows. The Fermi arc disperses systemtically into the unoccupied side as $k_y$ increases and becomes longer for higher Mo doping, consistent with topological theory, \textit{ab initio} calculation and the pump-probe ARPES spectra presented in the main text.}
\end{figure*}

\begin{figure*}[h]
\centering
\includegraphics[width=16cm,trim={1in 1in 1in 1in},clip]{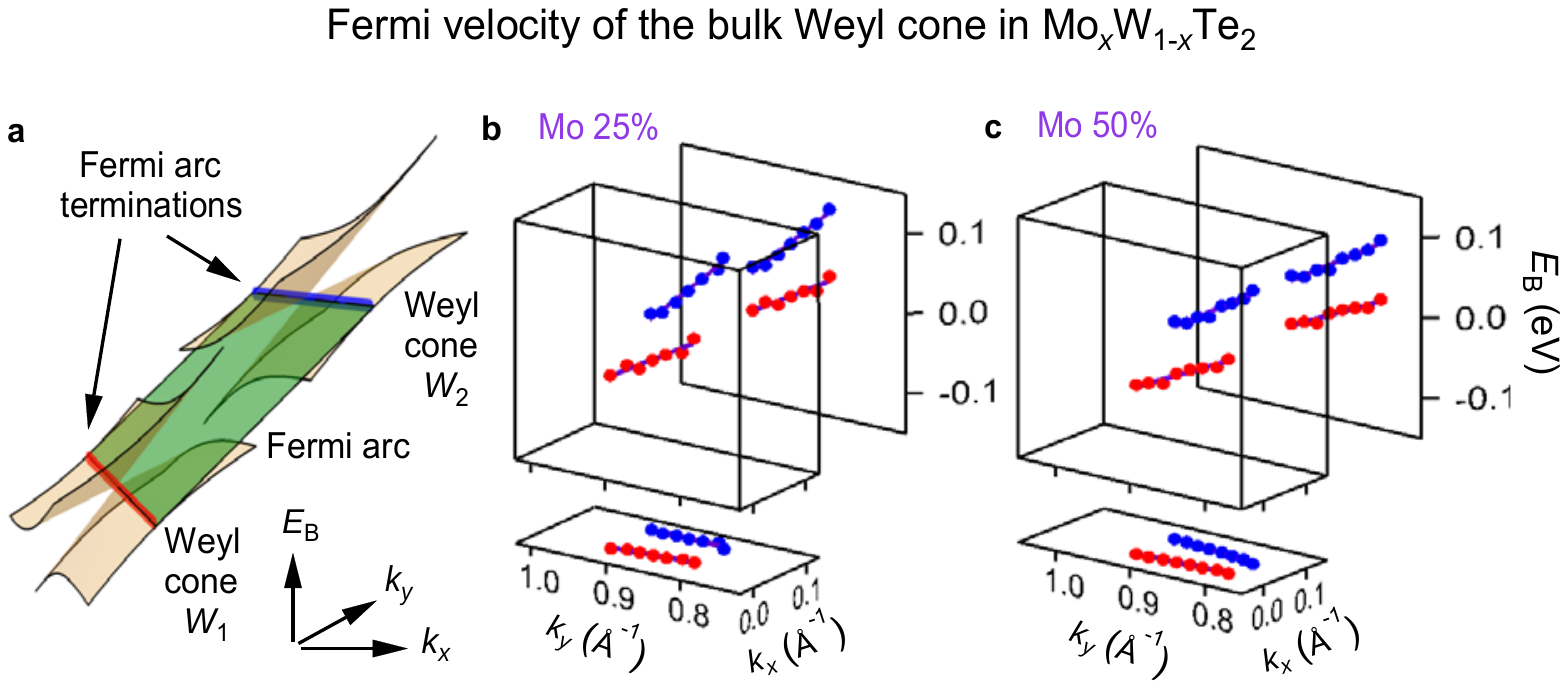}
\end{figure*}

\clearpage
\begin{figure*}
\centering
\caption{\label{Fig2} \textbf{Determining the bulk Weyl cone Fermi velocity from a surface state spectrum.} \textbf{a}, Illustration of the termination of the Fermi arc on the Weyl cones $W_1$ and $W_2$ in \mwt. By tracking the termination of the Fermi arc in momentum space (red and blue lines), we can measure the Fermi velocities of the Weyl cones along the attachment directions. \textbf{b}, \textbf{c}, Termination of the Fermi arc from pump-probe ARPES spectra, with a linear fit, for 20\% and 50\% Mo doping. We show the three-dimensional plot (box) and a projection into the $k_x$-$k_y$ plane (bottom projection) and $k_y$-$E_\textrm{B}$ plane (side projection). The numerical values of the fits are given in the text.}
\end{figure*}


\begin{thebibliography}{99}

\bibitem{QH} Prange, R. \& Girvin, R. \textit{The Quantum Hall Effect} (Springer, 1987).

\bibitem{TI_Suyang_2011} Xu, S.-Y. \textit{et al.} Topological Phase Transition and Texture Inversion in a Tunable Topological Insulator. \textit{Science} {\bf 332}, 560 (2011).

\bibitem{TaAsUs} Xu, S.-Y. \textit{et al}. Discovery of a Weyl fermion semimetal and topological Fermi arcs. \textit{Science} {\bf 349}, 613 (2015).
\bibitem{LingLu} Lu, L. \textit{et al}. Experimental observation of Weyl points. \textit{Science} {\bf 349}, 622 (2015).
\bibitem{TaAsThem} Lv, B. Q. \textit{et al}. Experimental discovery of Weyl semimetal TaAs. \textit{Phys. Rev. X} {\bf 5}, 031013 (2015).
\bibitem{TaAsThyUs} Huang, S.-M. \textit{et al}. A Weyl Fermion semimetal with surface Fermi arcs in the transition metal monopnictide TaAs class. \textit{Nat. Commun.} {\bf 6}, 7373 (2015).
\bibitem{TaAsThyThem} Weng, H. \textit{et al}. Weyl semimetal phase in noncentrosymmetric transition-metal monophosphides. \textit{Phys. Rev. X} {\bf 5}, 011029 (2015).
\bibitem{NbAs} Xu, S.-Y. \textit{et al.} Discovery of a Weyl fermion state with Fermi arcs in niobium arsenide \textit{Nat. Phys.} {\bf 11}, 748 (2015).
\bibitem{TaPUs} Xu, S.-Y. \textit{et al.} Experimental discovery of a topological Weyl semimetal state in TaP.  \textit{Sci. Adv.} {\bf 1}, 10 (2015).

\bibitem{Murakami} Murakami, S. Phase transition between the quantum spin Hall and insulator phases in 3D: emergence of a topological gapless phase. \textit{New Journal of Physics} {\bf 9}, 356 (2007).
\bibitem{Multilayer} Burkov, A. A. \& Balents, L. Weyl semimetal in a topological insulator multilayer. \textit{Phys. Rev. Lett.} {\bf 107}, 127205 (2011).
\bibitem{Pyrochlore} Wan, X. \textit{et al}. Topological semimetal and Fermi-arc surface states in the electronic structure of pyrochlore iridates. \textit{Phys. Rev. B} {\bf 83}, 205101 (2011).
\bibitem{Vish} Turner, A. \& Vishwanath, A. Beyond band insulators: topology of semi-metals and interacting phases. Preprint at http://arxiv.org/abs/1301.0330 (2013).

\bibitem{TR} Chang, T.-R. \textit{et al}. Prediction of an arc-tunable Weyl Fermion metallic state in \mwt. \textit{Nat. Commun.} {\bf 7}, 10639 (2016).
\bibitem{Andrei_WT_2015} Soluyanov, A.\textit{et al}. Type II Weyl semimetals. \textit{Nature} {\bf 527}, 495 (2015).
\bibitem{Zhijun_MT_2015} Wang, Z. J. \textit{et al}. MoTe$_2$: Weyl and line node topological metal. Preprint at http://arxiv.org/abs/1511.07440 (2015).
\bibitem{Binghai_MT_2015} Sun, Y. \textit{et al}. Prediction of the Weyl semimetal in orthorhombic MoTe$_2$. Preprint at http://arxiv.org/abs/1508.03501 (2015).


\bibitem{myothermowte} Belopolski, I. \textit{et al}. Fermi arc electronic structure and Chern numbers in the type-II Weyl semimetal candidate \mwt. \textit{Phys. Rev. B} {\bf 94}, 085127 (2016).
\bibitem{myotherothermowte} Belopolski, I. \textit{et al}. Discovery of a new type of topological Weyl fermion semimetal state in \mwt. \textit{Nat. Commun.} {\bf 7}, 13643 (2016).
\bibitem{HaoSTM} Zheng, H. \textit{et al}. Atomic-scale visualization of quasiparticle interference on a type-II Weyl semimetal surface. Preprint at https://arxiv.org/abs/1612.05208 (2016).

\bibitem{Adam1} Huang, L. \textit{et al}. Spectroscopic evidence for type II Weyl semimetallic state in MoTe$_2$. \textit{Nat. Mat.} DOI:10.1038/nmat4685 (2016).
\bibitem{Shuyun} Deng, K. \textit{et al}. Experimental observation of topological Fermi arcs in type-II Weyl semimetal MoTe$_2$. \textit{Nat. Phys.} DOI:10.1038/nphys3871 (2016).
\bibitem{Baumberger} Tamai, A. \textit{et al}. Fermi Arcs and Their Topological Character in the Candidate Type-II Weyl Semimetal MoTe$_2$. \textit{Phys. Rev. X} {\bf 6}, 031021 (2016).

\bibitem{MoTe2WTe2} Brown, B. E. The crystal structures of WTe$_2$ and high-temperature MoTe$_2$. \textit{Acta. Cryst.} {\bf 20} 268 (1966).

\bibitem{NbPme} Belopolski, I. \textit{et al}. Criteria for Directly Detecting Topological Fermi Arcs in Weyl Semimetals. \textit{Phys. Rev. Lett.} {\bf 116}, 066802 (2016).

\bibitem{IshidaMethods} Ishida, Y. \textit{et al}. Time-resolved photoemission apparatus achieving sub-20-meV energy resolution and high stability. \textit{Rev. Sci. Instr.} {\bf 85}, 123904 (2014).

\bibitem{GGA} Perdew, J. P., Burke, K. \& Ernzerhof, M. Generalized gradient approximation made simple. \textit{Phys. Rev. Lett.} {\bf 77}, 3865 (1996).
\bibitem{PAW1} Bl\"ochl, P. E. Projector augmented-wave method. \textit{Phys. Rev. B.} {\bf 50}, 17953 (1994).
\bibitem{PAW2} Kresse, G. \& Joubert, J. From ultrasoft pseudopotentials to the projector augmented-wave method. \textit{Phys. Rev. B.} {\bf 59}, 1758 (1999).
\bibitem{PlaneWaves1} Kresse, G. \& Furthm\"uller, J. Efficiency of \textit{ab initio} total energy calculations for metals and semiconductors using a plane-wave basis set. \textit{Comput. Mater. Sci.} {\bf 6}, 15 (1996).
\bibitem{Bonding} Mar, A., Jobic, S. \& Ibers, J. A. Metal-metal vs. tellurium-tellurium bonding in WTe$_2$ and its ternary variants TaIrTe$_4$ and NbIrTe$_4$. \textit{J. Am. Chem. Soc.} {\bf 114}, 8963 (1992).
\bibitem{MLWF1} Marzari, N. \& Vanderbilt, D. Maximally localized generalized Wannier functions for composite energy bands. \textit{Phys. Rev. B} {\bf 56}, 12847 (1997).
\bibitem{MLWF2} Souza, I., Marzari, N. \& Vanderbilt, D. Maximally localized Wannier functions for entangled energy bands. \textit{Phys. Rev. B} {\bf 65}, 035109 (2001).
\bibitem{Wannier90} Mostofi, A. A. \textit{et al}. Wannier90: a tool for obtaining maximally-localized Wannier functions. \textit{Comp. Phys. Commun.} {\bf 178}, 685 (2008).
\bibitem{MLWF3} Franchini, C. \textit{et al}. Maximally localized Wannier functions in LaMnO$_3$ within PBE$ + $U, hybrid functionals and partially self-consistent GW: an efficient route to construct \textit{ab initio} tight-binding parameters for $e_g$ perovskites. \textit{J. Phys. Cond. Mat.} {\bf 24}, 235602 (2012).
\bibitem{Green} Zhang, H. J. \textit{et al}. Topological insulators in Bi$_2$Se$_3$, Bi$_2$Te$_3$ and Sb$_2$Te$_3$ with a single Dirac cone on the surface. \textit{Nat. Phys.} {\bf 5}, 438 (2009).

\end{thebibliography}
\end{document}